\documentclass{jps-cp}
\usepackage{txfonts} 

\usepackage{amssymb}
\usepackage{amsmath}
\usepackage{amsfonts}
\usepackage{bm}
\usepackage{graphicx}
\usepackage{color}
\usepackage[usenames,dvipsnames]{xcolor}
\usepackage{dcolumn}
\usepackage{multirow}
\usepackage{array}
\usepackage{siunitx}

\usepackage{float}
\usepackage{physics}
\usepackage{threeparttable}

\newcommand{\ccax}{CeCu$_{1+x}$As$_2$}
\newcommand{\ccaeleven}{CeCu$_{1.11}$As$_2$}
\newif\iflogvar

\title{Anisotropic Physical Properties of the Kondo Semimetal CeCu$_{1.11}$As$_2$}

\author{Lukas \textsc{Cvitkovich}$^{1}$, Diego A. \textsc{Zocco}$^{1}$, Gaku \textsc{Eguchi}$^{1}$, Monika \textsc{Waas}$^{1}$, Robert \textsc{Svagera}$^{1}$, Berthold \textsc{Stöger}$^{2}$, Rajib \textsc{Mondal}$^{3}$, Arumugam \textsc{Thamizhavel}$^{3}$, and Silke \textsc{Paschen}$^{1}$}

\inst{$^{1}$Institute of Solid State Physics, Vienna University of Technology, 1040 Vienna, Austria \\ 
$^{2}$X-ray Center, Vienna University of Technology, 1040 Vienna, Austria \\ 
$^{3}$Department of Condensed Matter Physics and Materials Science, Tata Institute of Fundamental Research, Mumbai, India}

\email{zocco@ifp.tuwien.ac.at}

\recdate{September 17, 2019}

\abst{The recently proposed novel materials class called Weyl-Kondo semimetal (WKSM) is a time reversal invariant but inversion symmetry broken Kondo semimetal in which Weyl nodes are pushed to the Fermi level by the Kondo interaction. Here we explore whether \ccax\ may be a new WKSM candidate. We report on its single-crystal growth, structure determination and physical properties investigation. Previously published studies on polycrystalline samples suggest that it is indeed a Kondo semimetal, which is confirmed by our investigations on single crystals. X-ray diffraction reveals that \ccax\ crystallizes in a tetragonal centrosymmetric structure, although the inversion symmetry could still be broken locally due to partially occupied Cu sites. Chemical analysis results in an average occupation $x$\,=\,0.11(1). The electrical resistivity increases logarithmically with decreasing temperature, and saturates below 10\,K. A Kondo temperature $T_{\mathrm{K}}$\,$\approx$\,4\,K is extracted from entropy, estimated from the specific heat measurements. From Hall effect experiments, a charge carrier density of $\SI{8.8e20}{\per \cubic \centi \meter}$ is extracted, a value characteristic of a semimetal. The magnetization shows pronounced anisotropy, with no evidence of magnetic ordering down to 0.4\,K. We thus classify \ccaeleven\ as a tetragonal Kondo semimetal with anisotropic magnetic properties, with a possibly broken inversion symmetry, thus fulfilling the necessary conditions for a WKSM state.}

\kword{Kondo semimetal, Weyl-Kondo semimetal, Heavy fermion systems}

\begin{document}
\maketitle

\section{Introduction}

Kondo compounds are strongly correlated materials in which rare earth atoms with open 4$f$ shells are arranged in a regular lattice. Hybridization between localized states and the conduction band leads to the opening of a gap \cite{RiseboroughKI}. Depending on whether the Fermi energy is situated in the gap or within one of the hybridized bands, a Kondo insulator or a heavy fermion metal results. If the hybridization gap does not extend across the whole Brillouin zone, a Kondo semimetal is formed \cite{KSM_1, KSM_2}. Recently, a novel material class called Weyl-Kondo semimetal (WKSM) was proposed \cite{sami, WKSM_1, AHE2}, with Ce$_3$Bi$_4$Pd$_3$ as a prime candidate material. The formation of Weyl nodes in time reversal invariant materials requires that inversion symmetry is broken, as is indeed the case in cubic Ce$_3$Bi$_4$Pd$_3$.

In order to expand the catalog of WKSM candidates, we investigated the tetragonal compound \ccax\ ($x$\,$>$\,0). Previous studies on polycrystalline samples showed that \ccax\ crystallizes in a centrosymmetric tetragonal structure, with the space group $P$4/$nmm$, No. 129 \cite{WangCeCuAs2, CeCuAs_3}. The reason why we nevertheless consider it as a candidate WKSM is that it assumes a non-stoichiometric composition \cite{WangCeCuAs2} that may result in a symmetry reduction due to the excess of Cu atoms.

In this present work, we examine the composition, structure and physical properties of single crystalline samples of \ccax\ by energy dispersive x-ray spectroscopy, x-ray diffraction, electrical resistivity, Hall effect, magnetization and specific heat measurements. These experiments support the classification of \ccaeleven\ as a Kondo semimetal, and confirm the off-stoichiometric composition which allows for inversion symmetry being broken locally by partially occupied Cu sites.

\section{Materials and methods}\label{methods}

Single crystals of \ccax\ ($x$\,$>$\,0) were grown by a Cu-As self-flux method. From the binary phase diagram, Cu and As form an eutectic with a composition of 55:45 at.\,\% of Cu:As, melting at $\approx$\,600\,$^{\circ}$C. High purity Ce (Ames Lab, 99.9\%), Cu and As (Alfa Aesar, 99.9999\%) were used in the ratio of 1:11.26:10.74. The starting materials were taken in a high-quality recrystallized alumina crucible and subsequently sealed in a quartz ampoule with a partial pressure of argon gas. The ampoule was heated to 1050\,$^{\circ}$C at a rate of 15\,$^{\circ}$C/h in a box furnace and held at that temperature for 24\,h for a better homogenization of the solution. The furnace was then cooled to 845\,$^{\circ}$C at 1\,$^{\circ}$C/h, and the excess amount of flux was removed by centrifuging. As a result, thin shiny platelet single crystals of $\approx$\,2\,$\times$\,1\,$\times$ 0.1\,mm$^3$ were obtained.

Energy dispersive x-ray spectroscopy (EDX) measurements were performed at room temperature in high vacuum (\SI{5e-6}{\milli \bar}) using a scanning electron microscope with energy dispersive detector (EDAX New XL-30 135-10 UTW+). The energy of the excitation electrons was set to 30\,keV. 

Powder x-ray diffraction (XRD) measurements were performed in a PANalytical X'Pert Pro system with monochromatic Cu-K$_{\alpha}$ radiation. The samples were oriented at room temperature using Bruker and Huber Laue diffraction systems. Intensity data of single crystals were collected at 100\,K on a Bruker KAPPA Apex II diffractometer equipped with CCD detector (graphite monochromated Mo-K$_{\alpha}$ radiation, $\lambda$\,=\,0.71073\,\AA, 2$\theta_{\mathrm{max}}$\,=\,70$^{\circ}$). An initial model was generated using the published coordinates \cite{WangCeCuAs2}. The structure was refined against $F^2$ using Jana2006 \cite{Jana2006}.

Electrical contacts on the samples were made by spot-welding 25\,$\mu$m annealed gold wires. Electrical resistivity, Hall effect and magnetization measurements were performed between 300 and 2\,K in PPMS and MPMS systems from Quantum Design. The specific heat was also measured in the PPMS down to 0.4\,K, using the heat-pulse method. The samples were glued with Apiezon N-grease, and the addenda contribution was subtracted from the total heat capacity.

\section{Results and discussion}\label{results}

\subsection{Chemical and structural analysis}

\begin{figure}[t]
	\centering
	\includegraphics[width=1\linewidth]{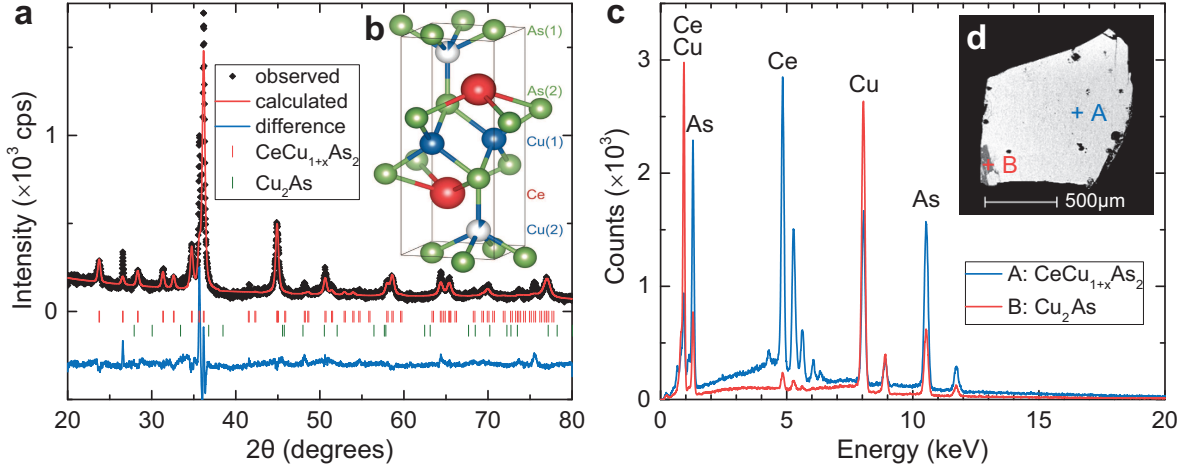}
	\caption{\textbf{a}: Room-temperature powder XRD of \ccax, consistent with the crystal structure sketched in \textbf{b}. Vertical bars indicate the 2$\theta$ positions of the expected diffraction peaks (red) and of the possible impurity phase Cu$_2$As (green). \textbf{c}: EDX spectra at two different spots A and B of a representative sample of the second batch. \textbf{d}: SEM image of the measured sample.}
	\label{fig1}
\end{figure}

Room temperature powder XRD confirmed the phase purity of \ccax, crystallizing in a stuffed variant of the HfCuSi$_2$-type tetragonal crystal structure (space group $P$4/$nmm$, No. 129) \cite{jemetio}, composed of puckered nets of Ce-As and Cu-As layers alternating along the [001] direction (Fig.~\ref{fig1}a and \ref{fig1}b). Laue diffraction revealed that the flat surfaces of the crystals correspond to the (001) planes. Due to the difficulty to obtain fine powders from the highly maleable single crystals, the refinement calculations did not match perfectly the intensities of all the observed peaks. However, we did not identify impurity phases from powder diffraction in the investigated sample, in particular of the expected Cu$_2$As phase (see below).

Initially, several samples from the first growth batch were screened by means of resistivity measurements from which a strong sample dependence was found, ranging from more insulating to more metallic. We suspect that the formation of metallic Cu$_2$As inclusions that crystallize in the same structure as \ccax\ were responsible for the variations \cite{Pauwels}. Indeed, initial EDX measurements on this first batch revealed that most of the examined samples contained metallic inclusions. When these inclusions were localized at the edge of the samples (far away from the electrical contacts), the metallic contribution to the transport measurements was negligible and thus the more insulating behavior prevailed. The foreign metallic phase was considerably reduced by adjusting the elements' ratios and the growth profile during a second crystal growth. Figure \ref{fig1}c displays the measured EDX spectra at two different spots of a representative sample (from the second growth batch, Fig.~\ref{fig1}d). The spectrum obtained for spot A results in the refined chemical composition \ccax\ with $x$\,=\,0.11. On the other hand, the region of different contrast located at the edge of the sample (spot B) is identified as the foreign phase Cu$_2$As. An average value of $x$\,=\,0.11(1) is obtained from EDX data collected from 5 samples of the second growth batch (Table~\ref{table1}, left). A crystal structure with a partially occupied Cu(2) site is consistent with our single-crystal XRD data and the corresponding structure refinement (Table~\ref{table1}, right), and in agreement with our EDX data as well as with the results obtained in previous studies \cite{WangCeCuAs2}. 

Following these results, and because previous measurements done on polycrystalline samples did not display metallic character \cite{WangCeCuAs2}, we identify the more insulating behavior as the intrinsic one. In the following sections, we report the physical properties of \ccaeleven\ samples from the second growth batch that are consistent with this intrinsic behavior.

\begin{table}[h]
	\centering
	\caption{
	Left: Averaged EDX data obtained from 11 measurements performed on 5 samples. Right: crystallographic data determined at 100\,K by single-crystal XRD. Lattice constants $a$\,=\,4.0396(4)\,$\AA$ and $c$\,=\,10.0708(16)\,$\AA$. Space group $P$4/$nmm$ No. 129, $Z$\,=\,2\,f.u./unit cell, density\,=\,7.3118\,g cm$^{-3}$. Number of reflections (collected, unique, observed): 1486, 261, 226. Residual factor R$_{\mathrm{obs}}$\,=\,3.6\,\%, wR$_{\mathrm{all}}$\,=\,9.2\,\%.}
	\begin{tabular}[t]{ccc}
		Atom	&		At.\,\%			&		per Ce	\\	\hline
		Ce-L	&		24.46(5)		&		1				\\
		Cu-K	&		27.22(8)		&		1.11(1)	\\
		As-K	&		48.32(9)		&		1.98(1)	\\
	\end{tabular}	
	\quad
	\quad
	\quad
	\begin{tabular}[t]{lcccc}
		Atom	&	Occup.	&		x			&		y		&		z					\\ \hline
		Ce		&		1			&		1/4		&	 1/4		&		0.24198(6) \\
		Cu(1)	&		1			&		3/4		&	 1/4		&		1/2				\\
		Cu(2)	&  0.12		&		3/4		&	 3/4		&		0.1133(12)	\\
		As(1)	&		1			&		3/4		&	 1/4		&		0					\\
		As(2)	&		1			&		1/4		&	 1/4		&		0.64936(12)
	\end{tabular}
	\label{table1}
\end{table}

\subsection{Electrical transport}
\begin{figure}[t]
	\centering
	\includegraphics[width=1\linewidth]{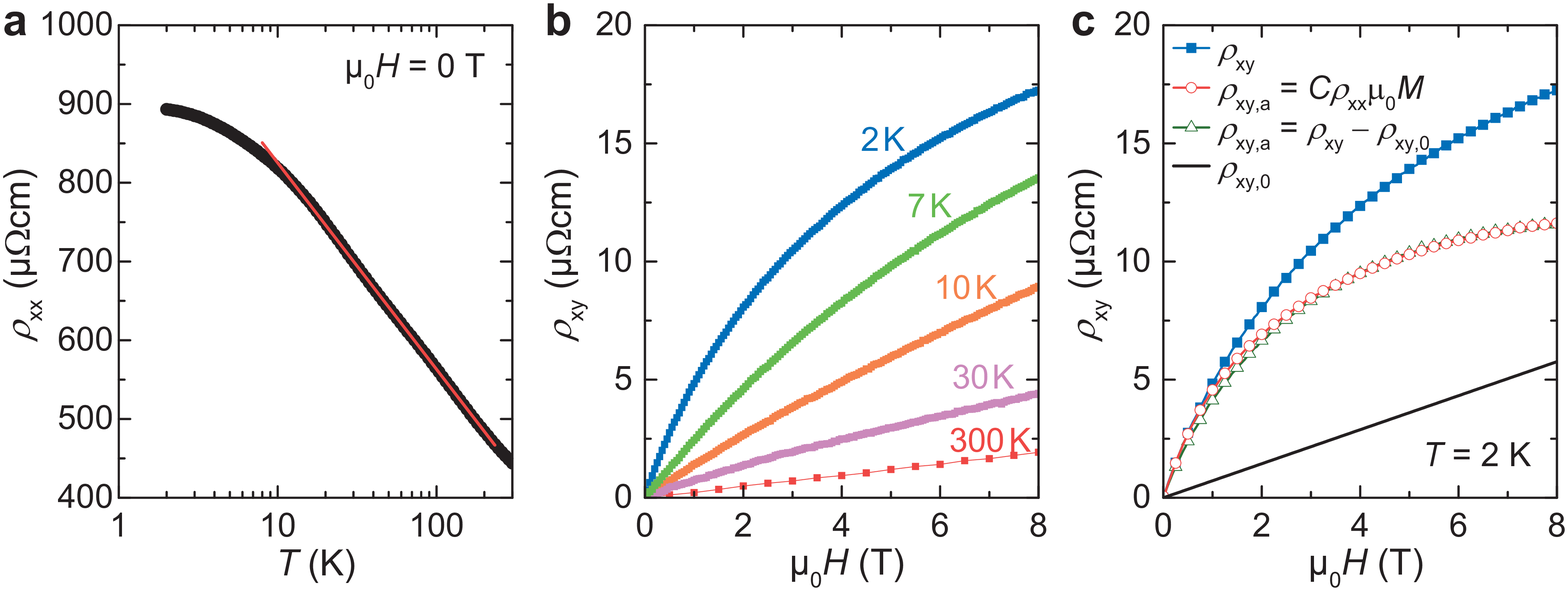}
	\caption{\textbf{Electrical transport measurements. a}: Resistivity \textit{vs.} temperature. The red line corresponds to a linear fit, indicating that $\rho_{xx}$\,$\propto$\,$-$ln($T$) above 20\,K. \textbf{b}: Hall resistivity $\rho_{xy}$ \textit{vs.} applied magnetic field at different temperatures. A pronounced nonlinearity develops at low temperatures. \textbf{c}: $\rho_{xy}$ measured at 2\,K (blue squares). It is the sum of an anomalous Hall resistivity $\rho_{xy,a}$ (green triangles), calculated as $\rho_{xy,a}$($H$)\,=\,$C \rho_{xx}$($H$)$\mu_0M$($H$) (red circles), and a linear Hall contribution $\rho_{xy,0}$ (black line).}
	\label{fig2}
\end{figure}

Measurements of the electrical resistivity reveal the canonical incoherent Kondo behavior at high temperatures, $\rho_{xx}$\,$\propto$\,$-$ln($T$) (Fig.~\ref{fig2}a). At temperatures below 10\,K the resistivity tends to saturate. This saturation is not found in heavy fermion metals, \textit{e.\,g.}\,CeAl$_3$ \cite{Fisk_1986, Ishida_CeAl3}, where the resistivity decreases rapidly with decreasing temperature after passing through a maximum. An inverse residual resistance ratio $iRRR$\,=\,$R$(2\,K)/$R$(300\,K)\,$\approx$\,2 was found for the most insulating samples.
\begin{figure}[b]
	\centering
	\includegraphics[width=1\linewidth]{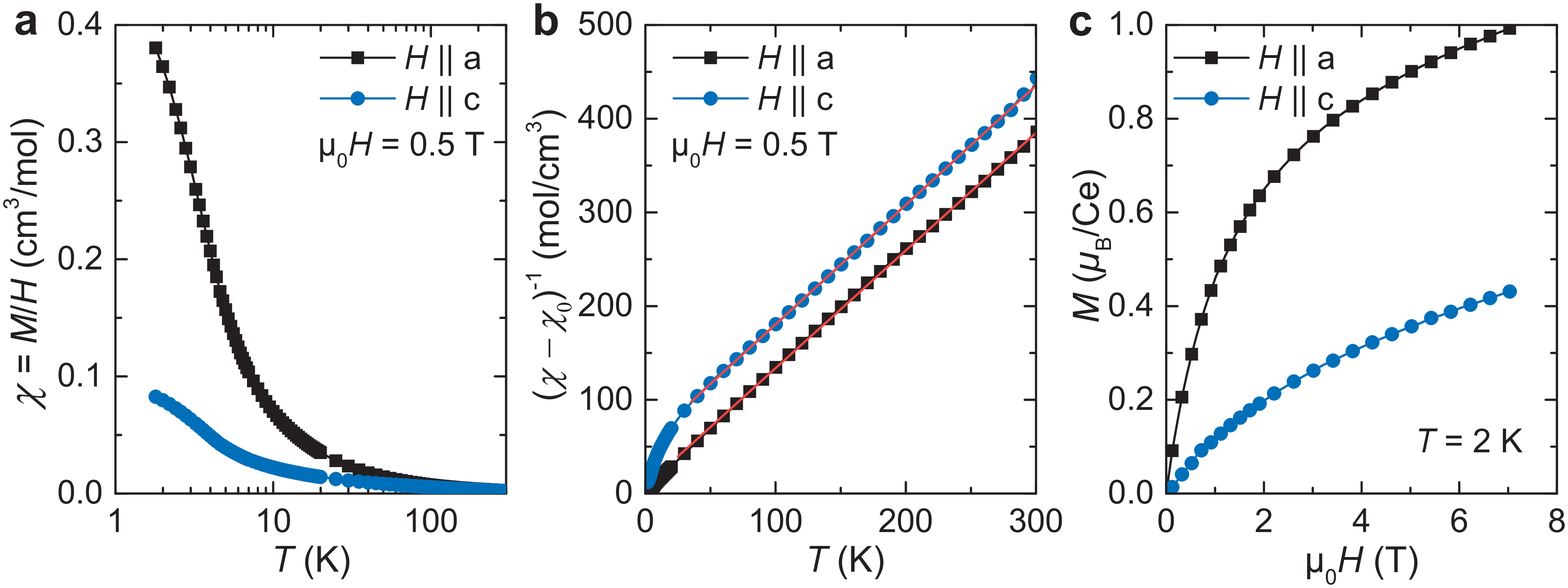}
	\caption{\textbf{Magnetization measurements. a}: Susceptibility \textit{vs.} temperature for magnetic fields applied along the $a$ and $c$ directions. \textbf{b}: Inverse susceptibility \textit{vs.} temperature. The red solid lines represent the Curie-Weiss fits. \textbf{c}: Magnetization per Ce atom \textit{vs.} applied magnetic field at $T$\,=\,2\,K.}
	\label{fig3}
\end{figure}

The Hall resistivity curves $\rho_{xy}$($H$) are displayed in Fig.~\ref{fig2}b; they develop strong nonlinearities at low temperatures. We identify this as an anomalous Hall effect (AHE) arising from the magnetization of the sample (see Section \ref{Magnetization}). In heavy fermion systems this AHE can be ascribed to skew scattering, and leads to an anomalous Hall resistivity $\rho_{xy,a}$($H$)\,=\,$C \rho_{xx}$($H$)$\mu_0 M$($H$) \cite{AHE_paper}, where the resistivity $\rho_{xx}$ and the magnetization $M$ are measured with $H$\,$\parallel$\,$z$ (in our case, $z$\,$\equiv$\,$c$) and $C$ is a scaling factor. The total Hall resistivity corresponds to the sum of a linear-in-field normal Hall contribution $\rho_{xy,0}$ and the AHE term, \textit{i.\,e.}\,$\rho_{xy}$\,=\,$\rho_{xy,0}$\,+\,$\rho_{xy,a}$, which are plotted in Fig.~\ref{fig2}c. A charge carrier concentration $n=\SI{8.8e20}{\per \cubic \centi \meter}$ is obtained from $\rho_{xy,0}$ at 2\,K, which is typical for a semimetal.

\subsection{Magnetization}\label{Magnetization}

The magnetic susceptibility displays paramagnetic behavior and a pronounced anisotropy with a weaker paramagnetic response for $H$\,$\parallel$\,$c$ (Fig.~\ref{fig3}). No evidence of magnetic ordering was found down to 1.8\,K. At high temperatures the susceptibilities follow a modified Curie-Weiss law $\chi$($T$) = $\chi_{\mathrm{0}}$\,+$C$/($T-\theta_{\mathrm{P}}$), where $\chi_{\mathrm{0}}$ is the temperature independent part of the susceptibility (mainly from core-electron diamagnetism and conduction-electron Pauli susceptibility) and $C$ is the Curie constant, with $\chi_{\mathrm{0}}$\,=\,$-$7.72\,$\times$\,10$^{-5}$ cm$^3$/mol, $\mu_{\mathrm{eff}}$\,=\,2.54$\mu_{\mathrm{B}}$ and $\theta_{\mathrm{P}}$\,=\,$-$8.98\,K for $H$\,$\parallel$\,$a$, and $\chi_{\mathrm{0}}$\,=\,$-$2.28\,$\times$\,10$^{-4}$ cm$^3$/mol, $\mu_{\mathrm{eff}}$\,=\,2.51$\mu_{\mathrm{B}}$ and $\theta_{\mathrm{P}}$\,=\,$-$42.01\,K for $H$\,$\parallel$\,$c$. The obtained effective moments are in good agreement with the value expected for a free Ce$^{3+}$ ion, \textit{i.\,e.}  $\mu_{\mathrm{eff}}^{\mathrm{Ce}^{3+}}$\,=\,2.54$\mu_\mathrm{B}$ for a Hund's rule ground state with $J$\,=\,5/2. 

\subsection{Specific heat}
\begin{figure}[t]
	\centering
	\includegraphics[width=1\linewidth]{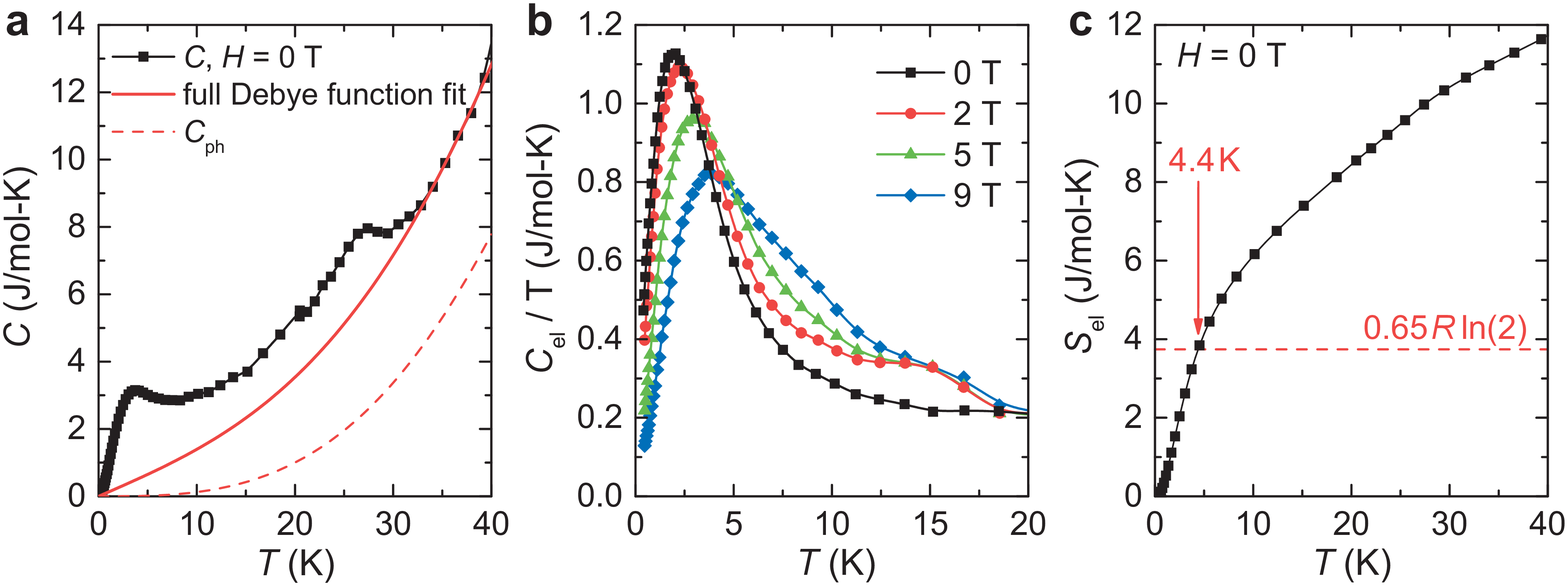}
	\caption{\textbf{Specific heat measurements. a}: Total specific heat at $H$\,=\,0. The red solid and dashed lines correspond to the phonon background analysis. \textbf{b}: Electronic specific heat at 0, 2, 5, and 9\,T. \textbf{c}: Electronic contribution to the entropy $S_{\mathrm{el}}$ \textit{vs.} $T$. A Kondo temperature $T_{\mathrm{K}}$\,$\approx$\,4.4\,K is obtained when $S_{\mathrm{el}}$ reaches 0.65Rln(2) \cite{degrange}.}
	\label{fig4}
\end{figure}

The specific heat of \ccaeleven\ was measured above 0.4\,K, and is plotted in Fig.~\ref{fig4}a. Two broad features are observed slightly below 30\,K and 10\,K, although neither of them can be identified as a phase transition. In order to obtain the electronic contribution to the specific heat, the phonon background was extracted by fitting the data above 30\,K using the full Debye-Sommerfeld model \cite{gaku}, which yields a Debye temperature $\theta_{\mathrm{D}}$\,=\,(323\,$\pm$\,42)\,K.  After subtracting the phononic contribution (dashed line in Fig.~\ref{fig4}a), we obtained the electronic contribution to the specific heat, plotted in Fig.~\ref{fig4}b for different magnetic fields. The broad feature at low temperatures becomes a prominent rounded peak centered at 2\,K, which shifts towards higher temperatures for increasing applied magnetic fields. Similar broad features were previously observed in polycrystalline samples \cite{CeCuAs_3}, although further experiments and analysis are required to clarify their origin.

The electronic contribution to the entropy $S_{\mathrm{el}}$ is obtained by integrating $C_{\mathrm{el}}/T$\,\textit{vs.}\,$T$, and it is displayed in Fig.~\ref{fig4}c. The Kondo temperature $T_{\mathrm{K}}$ can be estimated as $S_{\mathrm{el}}$($T_{\mathrm{K}}$)\,=\,0.65ln(2) \cite{degrange}, which results in $T_{\mathrm{K}}$\,$\approx$\,4.4\,K.

\section{Summary and conclusions}
We have successfully grown single crystals of \ccax\ by the self-flux method. The results of the powder and single-crystal x-ray diffraction as well as of the energy-dispersive x-ray spectroscopy experiments are consistent with a tetragonal, centrosymmetric structure, and a partially occupied Cu(2) atomic site with $x$\,=\,0.11(1) that may result in a reduction of the crystal symmetry. A pronounced anisotropy is observed in magnetization, and no evidence of magnetic ordering is found down to 0.4\,K. Electrical resistivity and Hall measurements are consistent with a description of \ccaeleven\ as a Kondo semimetal. We hope this work will trigger further explorations of \ccaeleven\ and other Ce-based Kondo semimetals as possible candidates to host exotic Weyl-Kondo semimetallic states.

\section*{Acknowledgements}
The work at the Technical University in Vienna was financially supported by the Austrian Science Fund (FWF projects I2535-N27 and I4047-N27). We thank X. Yan, A. Prokofiev, M. Taupin and S. Dzsaber for helpful discussions.

\end{document}